\documentclass[aps,prc,reprint,eqsecnum,showpacs,floatfix]{revtex4-1}

\usepackage{graphicx,epstopdf}
\usepackage{hyperref}
\usepackage{color}
\usepackage{subfigure}
\usepackage{amsmath}
\usepackage{amssymb}
\usepackage{epsfig}
\usepackage{amsfonts}
\usepackage{mathtools}
\usepackage{bbm}
\usepackage{bm}
\usepackage{float}
\usepackage{xcolor}

\def\rvec{{\bf r}}
\def\hvec{{\bf h}}
\def\kvec{{\bf k}}
\def\pvec{{\bf p}}
\def\qvec{{\bf q}}

\def\bra#1{\left\langle#1\right|}
\def\ket#1{\left|#1\right\rangle}

\def\Im{{\cal I}m}

\def\Tr{{\cal T}r}
\def\KF{k_{\rm F}}
\def\SF{S_{\rm F}}
\def\tF{t_{\rm F}}
\def\1{\mathbbm{1}}

\def\bra#1{\bigl\langle{ #1} \bigr|}
\def\ket#1{\bigl|{ #1} \bigr\rangle}
\def\Bra#1{\Bigl\langle{ #1} \Bigr|}

\def\ovlp#1#2{\bigl\langle{ #1}\big|{#2} \bigr\rangle}
\def\Ovlp#1#2{\Bigl\langle{ #1}\Big|{#2} \Bigr\rangle}

\def\ie{{\em i.e.\/}\ }

\def\sij#1#2{{\bm\sigma}_{#1}\cdot{\bm\sigma}_{#2}}
\def\srij#1#2{\left({\bm\sigma}_{#1}\cdot\hat\rvec\right)
  \left({\bm\sigma}_{#2}\cdot\hat\rvec\right)}

\def\boxit#1{
        \centerline{\vbox{\hsize=6.0truein\hrule\hbox{\vrule\kern5pt
        \vbox{\kern5pt\noindent #1\smallskip
        \kern5pt}\kern5pt\vrule}\hrule}
}}

\begin{document}

\title{Variational and parquet-diagram calculations for neutron
  matter.\\ II.  Twisted Chain Diagrams.}

\author{E.~Krotscheck and J. Wang}

\affiliation{Department of Physics, University at Buffalo, SUNY
Buffalo NY 14260}
\affiliation{Institut f\"ur Theoretische Physik, Johannes
Kepler Universit\"at, A 4040 Linz, Austria}

\begin{abstract}
  We develop a manifestly microscopic method to deal with strongly
  interacting nuclear systems that have different interactions in
  spin-singlet and spin-triplet states. In a first step we analyze
  variational wave functions that have been suggested to describe such
  systems, and demonstrate that the so-called commutator contributions
  can have important effects whenever the interactions in the
  spin-singlet and the spin-triplet states are very different. We then
  identify these contributions as terms that correspond, in the
  language of perturbation theory, to non-parquet diagrams. We include
  these diagrams in a way that is suggested by the Jastrow-Feenberg
  approach and show that the corrections from non-parquet
  contributions are, at short distances, larger than all other
  many-body effects.
\end{abstract}

\maketitle

\section{Introduction}

Popular models of the nucleon-nucleon forces
\cite{Reid68,Bethe74,Day81,AV18,Wiri84} represent the interaction as a
sum of local functions times correlation operators, \ie
\begin{equation}
\hat v (i,j) = \sum_{\alpha=1}^n v_\alpha(r_{ij})\,
        \hat O_\alpha(i,j),
\label{eq:vop}
\end{equation}
where $r_{ij}=|\rvec_i-\rvec_j|$ is the distance between particles $i$
and $j$, and the $O_\alpha(i,j)$ are operators acting on the spin,
isospin, and possibly the relative angular momentum variables of the
individual particles.  According to the number of operators $n$, the
potential model is referred to as a $\hat v_n$ model
potential. Semi-realistic models for nuclear matter keep at least the
six base operators, these are
\begin{eqnarray}
\hat O_1(i,j;\hat\rvec_{ij})
        &\equiv& \hat O_c = \1\,,
\nonumber\\
\hat O_3(i,j;\hat\rvec_{ij})
        &\equiv& {\bm\sigma}_i \cdot {\bm\sigma}_j\,,
\nonumber\\
\hat O_5(i,j;\hat\rvec_{ij})
&\equiv& S_{ij}(\hat\rvec_{ij})
      \equiv 3({\bm\sigma}_i\cdot \hat\rvec_{ij})
      ({\bm\sigma}_j\cdot \hat\rvec_{ij})-{\bm\sigma}_i \cdot {\bm\sigma}_j\,,
      \nonumber\\
      \hat O_{2n}(i,j;\hat\rvec_{ij}) &=& \hat O_{2n-1}(i,j;\hat\rvec_{ij})
      {\bm\tau}_1\cdot{\bm\tau}_2\,.
\label{eq:operator_v6}
\end{eqnarray}
where $\hat\rvec_{ij} = \rvec_{ij}/r_{ij}$. We will omit the
arguments when unambiguous.

Besides the operators defined in Eq. \eqref{eq:operator_v6}
it is convenient to introduce the projection operators
\begin{align}
  \hat P_{S\phantom{+}} &\equiv \frac{1}{4}\left(\1-{\bm\sigma}_1\cdot{\bm\sigma}_2\right)\,,
  \nonumber\\
  \hat P_{T+} &\equiv \frac{1}{6}\left(3\1+\sij12+S_{12}(\hat\rvec)
  \right)\,,\label{eq:projectors}\\
  \hat P_{T-} &\equiv \frac{1}{12}\left(3\1+\sij12-2S_{12}(\hat\rvec)
  \right)\,.
  \nonumber
  \end{align}
These satisfy the relations $\hat P_i \hat P_j = \hat P_i\delta_{ij}$ and
$\hat P_S + \hat P_{T+} + \hat P_{T-} = \1$.

If tensor forces are included, a third set of operators
\begin{equation}
  \hat L \equiv \srij12\,,
  \quad
  \hat T \equiv \sij12-\srij12\,.\label{eq:CLTbasis}
\end{equation}
 is  useful for the summation of chain diagrams
\cite{FNN82}.

The task of microscopic many-body theory is to understand properties
of macroscopic systems from no other information than the properties
of the underlying Hamiltonian, the particle statistics, and the
macroscopic geometry of the system. For simple, state-independent
interactions as appropriate for electrons or quantum fluids, the
Jastrow-Feenberg ansatz \cite{FeenbergBook} for the wave function
\begin{equation}
\Psi_0 = \prod^N_{i,j=1 \atop i<j} f(r_{ij})\Phi_0
\label{eq:Jastrow}
\end{equation}
and its logical generalization to multiparticle correlation functions
\cite{Woo3body1,Woo3body2,Chuckphonon,EKthree,polish} has been
extremely successful. $\Phi_0$ is, for fermions, a Slater determinant
of single particle orbitals. The method has therefore been applied in
both semi-analytic calculations \cite{FeenbergBook} as well as early
Monte Carlo calculations \cite{KalosLevVer,CeperleyVMC} and is still
being used as an importance sampling function for diffusion and
Green's functions Monte Carlo computations
\cite{CeperleyRMP,JordiQFSBook}.

One of the reasons for the success of the wave function
\eqref{eq:Jastrow} is that it provides a good upper bound for the
ground state energy
\begin{equation}
E_0 = \frac{\bra{\Psi_0}H\ket{\Psi_0}}{\ovlp{\Psi_0}{\Psi_0}}\,.
  \label{eq:energy}
\end{equation}
In semi-analytic calculations, approximations must be made in the
evaluation of the energy expectation value \eqref{eq:energy}.  The
hierarchy of ``(Fermi-)hypernetted chain ((F)HNC)'' approximations
\cite{LGB} is singled out since it permits an unconstrained
optimization of the correlation functions,
\begin{equation}
\frac{\delta E_0}{\delta f}(\rvec_1,\rvec_2) = 0,
\label{eq:euler}
\end{equation}
in the sense that the Euler equations for any level of the HNC
approximation has the same structure as the exact Euler equation
\cite{FeenbergBook}.  The method corresponds, for bosons, to a
self-consistent summation of all ring and ladder diagrams of
perturbation theory -- the so-called ``parquet'' diagrams
\cite{Woo70,parquet1,parquet2,parquet3}. The same is true for Fermions
\cite{EKvar} when specific truncation orders of exchange diagrams are
followed.

The Jastrow-Feenberg ansatz (\ref{eq:Jastrow}) is insufficient for
dealing with realistic nucleon-nucleon interactions of the form
(\ref{eq:vop}). A plausible generalization of the wave function
(\ref{eq:Jastrow}) is the symmetrized operator product
\cite{FantoniSpins,IndianSpins}
\begin{equation}
        \Psi_0^{{\rm SOP}}
        = {\cal S} \Bigl[ \prod^N_{i,j=1 \atop i<j} \hat f (i,j)\Bigr] \Phi_0\,,
\label{eq:f_prodwave}
\end{equation}
where
\begin{equation}
  \hat f(i,j) = \sum_{\alpha=1}^n f_\alpha(r_{ij})\,
  \hat O_\alpha(i,j)\,,
  \label{eq:fop}
\end{equation}
and ${\cal S}$ stands for symmetrization. The symmetrization is
necessary because the operators $\hat O_\alpha(i,j)$ and $\hat
O_\beta(i,k)$ do not necessarily commute. The potential energy, for
example, can be written in the form
\begin{equation}
\frac{\left\langle V \right\rangle}{N}
= \frac{\rho}{2}\int d^3r \sum_\alpha g_\alpha(r) v_\alpha(r)
\frac{1}{\nu^2}\Tr_{12}\, O_\alpha^2(1,2)
\label{eq:epot}
\end{equation}
where $\nu$ is the degree of degeneracy of the single-particle states,
$\Tr_{12}$ indicates the trace over spin (and, when applicabe,
isospin) variables of particles 1 and 2, and the
components of the pair distribution function are
\begin{align}
  &\rho^2 g_\alpha(|\rvec-\rvec'|)=\nonumber\\
  &
  \frac{\Bra{\Psi_0}\sum_{i<j}\delta(\rvec-\rvec_i)\delta(\rvec'-\rvec_j)
    \hat O_\alpha(i,j)\ket{\Psi_0}}{\displaystyle\frac{1}{\nu^2}\Tr_{12}\, \hat O_\alpha^2(1,2)
    \Ovlp{\Psi_0}{\Psi_0}}\,.\label{eq:gop}
\end{align}

The need to symmetrize the operator product causes, however, severe
problems which must be dealt with properly: When the symmetrization is
carried out, the components of the pair distribution operator have the
form
\begin{equation}
g_\alpha(r) =\sum_{\beta\gamma}
f_\beta(r)
f_\gamma(r)F^{(\alpha)}_{\beta\gamma}(r)\label{eq:gsymmetrized}
\end{equation}
where the $F^{(\alpha)}_{\beta\gamma}(r)$ are coupling coefficients that
are functionals of the correlation functions $f_\alpha(r_{ij})$.
Their analytic structure is complicated and so far no summation that
comes anywhere close to the diagrammatic richness of the (F)HNC
summations for state-independent correlations has been found.

The only relevant feature for our analysis is, however, that the
coefficient functions $F^{(\alpha)}_{\beta\gamma}(r)$ are {\em not\/}
diagonal in the operator labels $\alpha$, $\beta$, and
$\gamma$. In other words, the interaction in operator channel $\alpha$
is, in the potential energy, multiplied with correlation functions
$f_\beta(r)f_\gamma(r)$ with $\beta\ne\alpha$, $\gamma\ne\alpha$.

This is {\em a priori\/} not a problem because the (observable) pair
distribution functions $g_\alpha(r)$ can be thought of as the
independent quantities in the variational problem, \ie instead of
Eq. \eqref{eq:euler} we may use
\begin{equation}
\frac{\delta E_0}{\delta g_\alpha}(\rvec_1,\rvec_2) = 0\,.
\label{eq:eulerG}
\end{equation}
Then, the theory can be formulated entirely in terms of observable
quantities. In fact the basic equations of the boson theory can be
derived in several ways
\cite{parquet1,parquet2,PairDFT,BishopValencia} without ever
mentioning the auxiliary Jastrow correlation function $f(r)$.

\begin{figure}
  \centerline{\includegraphics[width=0.7\columnwidth,angle=270]
    {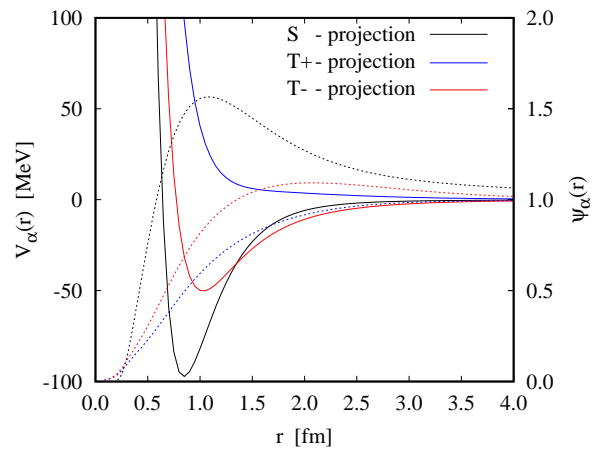}}
  \caption{The Reid interaction \cite{Reid68} in the $\{S,T+,T-\}$
    operator form for the singlet (black line), Triplet ``+'' (blue
    line) and Triplet ``-'' (red line) projections (left scale). Also
    shown are the corresponding pair wave functions $\psi_\alpha(r) =
    \sqrt{1+\Gamma_{\rm dd}^{(\alpha)}(r)}$, see Eq. \eqref{eq:psiGdd}
    at $\KF = 1.0\,\mathrm{fm}^{-1}$ (dashed line, same colors, right
    scale).}\label{fig:Reid}
\end{figure}

However, if one adopts the original idea of Jastrow theory and uses
some parameterized form of the correlation functions $f_{\alpha}(r)$,
a good parametrization is hard to find.  A popular choice
\cite{ScottMozkowski,PandharipandeBethe} is, for example, to derive
the correlation functions $f_{\alpha}(r)$ from a low-order constrained
variational principle (LOCV).  This leads for the correlation
functions to an effective Schr\"odinger equation of the form
\begin{align}
  -&\frac{\hbar^2}{m}\nabla\cdot\left[g_F^{(\alpha)}(r)\nabla f_{\alpha}(r)\right]
  \nonumber\\
      +&(v_{\alpha}(r) - \lambda_{\alpha})g_{\rm F}^{(\alpha)}(r) f_{\alpha}(r)=0
  \label{eq:LOCV}\,.
\end{align}
Eq. \eqref{eq:LOCV} is understood in the projection operator basis
\eqref{eq:projectors}.  The $\lambda_{\alpha}$ are parameters
determined either by a normalization condition
\cite{ScottMozkowski,OBI76} or by the demand that $f_\alpha'(d) = 0$
at a healing distance $d$ \cite{PandharipandeBethe}, and
$g_F^{(\alpha)}(r) = 1 \pm \ell^2(r\KF)$ are the distribution
functions on non-interacting fermions, the upper/lower sign refers to
singlet/triplet states, and $\ell(x)=3j_1(x)/x$.  Modern
nucleon-nucleon interactions \cite{Reid68,Bethe74,Day81,AV18,Wiri84}
have rather different core sizes in the spin-singlet and the
spin-triplet cases, see for example Fig. \ref{fig:Reid}

The above analysis shows that the commutator terms in the symmetrized
operator product \eqref{eq:f_prodwave} mix different channels such
that it is by no means clear how well the short-ranged correlations
are described by simple approximations like \eqref{eq:LOCV}. In fact,
earlier nucleon-nucleon interaction \cite{MozkowskiScott,Ohmura69}
contained hard cores with different core sizes in different operator
channels. In that case, the potential energy obtained from
correlations determined by the LOCV method diverges already if only the
simplest non-trivial commutator term is retained.

As an alternative to variational wave functions, Smith and Jackson
\cite{SmithSpin} started from the idea of localized parquet-diagram
summations and implemented the procedure for a fictive system of
bosonic nucleons interacting via a $v_6$ interaction. It turned out
that the equations derived were identical to those one would obtain in
the bosonic version of the summation method of Fantoni and Rosati
\cite{FantoniSpins}, which simply ignored the fact that the individual
pair correlation operators $\hat f(i,j)$ do not commute. We have
adopted in Ref. \onlinecite{v3eos} the ideas of Smith and Jackson and
generalized them to Fermi systems. In that work, we have also paid
attention to different treatments of the particle-particle and the
particle-hole propagator and have determined which approximations for
these quantities are suggested by variational wave functions.

The problem of the potential importance of commutator diagrams does
not go away in parquet summations. The fact that the work of
Ref. \onlinecite{SmithSpin} corresponds to a variational calculation
where all commutators are omitted simply says that the fully
symmetrized wave function \eqref{eq:f_prodwave} contains more than
what is included in the parquet equations. The analysis
\eqref{eq:gsymmetrized} shows that these non-parquet contributions are
important. On the other hand, the point of view of parquet-diagram
summations promises a more straightforward procedure to deal with
these effects compared to going through the development of a full
variational procedure. The equivalence between classes of Feynman
diagrams and classes of Jastrow-Feenberg diagrams will provide a
vehicle for justifying practical procedures for their calculation. To
that end, we will in the next section discuss, with a very simple
example, how the physics of commutator corrections is described in
terms of Goldstone diagrams and which approximations to these diagrams
are suggested by a variational wave function.

Section \ref{ssec:twist} will then derive the implementation of these
``non-parquet'' diagrams in a generalized Bethe-Goldstone equation.
Numerical applications will be discussed in Section \ref{sec:results},
Section \ref{sec:summary} will provide a brief summary of our
findings. The appendix will review earlier work \cite{SpinTwist} where
the symmetrization problem can be examined in a relatively simple
analytic form.

\section{Essentials of  parquet diagram
  summations and the optimized hypernetted chain method}
\label{sec:FHNCEL}

The basic insight, which was explained quite convincingly in
Ref. \onlinecite{parquet1}, is that the minimal satisfactory
microscopic treatment of an interacting system of many identical
particles requires the self-consistent summation of ring- and ladder
diagrams. This is in principle an exceedingly demanding task because
each two body vertex is a functions of two incoming
$(\kvec_i,\hbar\omega_i)$ and two outgoing sets of four quantum
numbers. Energy and momentum conservation as well as isotropy lets us
reduce the number of variables to 10. Hence, approximations must be
made to make the theory practical which we review here.

\subsection{Ring diagrams and the induced interaction}

The ring diagrams describe low-lying excitations and long-ranged
correlations. The sum of ring diagrams diverges when the system is
unstable against low-lying excitations such as density- or
spin-density fluctuations. Therefore, their inclusion is important to
have the correct non-analytic density dependence of the equation of
state of a self-bound system.

These are derived from a random-phase
approximation (RPA) equation for the response function
 \begin{align}
    \hat \chi(q,\omega) &=
    \frac{\chi_0(q,\omega)} {1-\hat V_{\rm
        p-h}(q)\chi_0(q,\omega)}\nonumber\\
 S(q) &= -\int_0^\infty \frac{d\hbar\omega}{\pi} \Im \chi(q,\omega),
\label{eq:SRPA}
 \end{align}
in terms of a local ``particle-hole'' interaction $\hat V_{\rm
  p-h}(q)$. In the case of state-dependent interactions, $\hat V_{\rm
  p-h}(q)$ is most conveniently represented as a linear combination of
local functions $\tilde V^{(\alpha)}_{\rm p-h}(q)$ and the operators
\eqref{eq:CLTbasis}.  As usual we define the dimensionless Fourier
transform by including a density factor $\rho$, \ie
  \begin{equation}
    \tilde f(k) = \rho\int d^3r f(r) e^{i{\bf k}\cdot\rvec}
    = \rho\int d^3r f(r) j_0(kr)\,.\label{eq:ft}
  \end{equation}
For the tensor forces, we will also need the $j_2$ Fourier transform,
    \begin{eqnarray}
      \tilde f(k)S_{12}(\hat\kvec)
      &=& \rho\int d^3r f(r) e^{i{\bf k}\cdot\rvec}S_{12}(\hat\rvec)\nonumber\\
    &=& -\rho\int d^3r f(r) j_2(kr)S_{12}(\hat\kvec)\,.\label{eq:ft2}
  \end{eqnarray}

The second important relationship is the Bethe-Goldstone equation
which describes short-ranged correlations caused by the strong,
short-ranged repulsion of the nuclear interaction. We shall discuss
this in detail below.  Summing the parquet diagrams one supplements,
among others, the bare interaction $\hat v(\rvec)$ in the Bethe-Goldstone
equation by an induced interaction $\hat w_{\rm I}(\rvec)$ being defined as the
set of particle-hole reducible diagrams. Assuming a particle-hole
interaction $\hat V_{\rm p-h}(\qvec)$ of the operator of the form
\eqref{eq:operator_v6}, the sum of these diagrams is {\em a priori} an
energy-dependent quantity
\begin{equation}
  \hat w_{\rm I}(\qvec,\omega) =  \frac{\hat V_{\rm p-h}^2(\qvec)\chi_0(q,\omega)}
  {1-\hat V_{\rm p-h}(\qvec)\chi_0(q,\omega)}\,.
\end{equation}
The energy dependent induced interaction is then approximated
\cite{parquet1,parquet2} by an energy independent effective
interaction $\hat w(q)$ as follows: Calculate the static structure
function
\begin{align}
  S(\qvec) &= - \int_0^\infty \frac{d\hbar\omega}{\pi}\,
  \Im \frac{\chi_0(q,\omega)}{1- \chi_0(q,\omega)\hat V_{\rm p-h}(\qvec)}
  \nonumber\\
  & = - \int_0^\infty \frac{d\hbar\omega}{\pi}\,
  \Im \left[\chi_0(q,\omega) + \chi_0^2(q,\omega)\hat w_{\rm I}(\qvec,\omega)\right]\,.
\end{align}
Now define an {\em energy independent interaction\/} $\hat
w_{\rm I}(q,\bar\omega(q))$ by demanding that it gives the same static
structure function,
\begin{equation}
  S(q)\equiv -\int_0^\infty \frac{d\hbar\omega}{\pi}\,
  \Im \left[\chi_0(q,\omega) + \chi_0^2(q,\omega)\hat w_{\rm I}(q,\bar\omega(q))\right]\,.
\label{eq:Scond}
\end{equation}

This energy independent interaction $\hat w_{\rm I}(\qvec) \equiv \hat
w_{\rm I}(\qvec,\bar\omega(q))$ is then taken as a correction to the interaction
in the Bethe-Goldstone equation. For state-dependent interactions it
is again understood that $\hat w_{\rm I}(\qvec,\omega)$ is a linear combination
of local functions and operators of any of the forms
\eqref{eq:operator_v6}, \eqref{eq:projectors} or \eqref{eq:CLTbasis}.

\subsection{Localized Bethe-Goldstone equation}
\label{sec:BG}

We review here briefly the connection between the conventional
Bethe-Goldstone equation and the variational approach.  We begin with
the Bethe-Goldstone equation as formulated in Eqs.~(2.1), (2.2) of
Ref.~\onlinecite{BetheGoldstone57}. As a convention, we will label
occupied (``hole'') states by $\hvec, \hvec', \hvec_i$ and unoccupied
(``particle'') states by $\pvec,\pvec',\pvec_i$; whereas $\kvec$,
$\qvec$ have no restriction. We also suppress spin variables. The pair
wave function $\psi(r)$ in a coordinate frame centered at the origin of the Fermi sea is given by the
integral equation
\begin{align}
  \bra{\kvec,\kvec'}\psi\ket{\hvec,\hvec'}
  &=  \ovlp{\kvec,\kvec'}{\hvec,\hvec'}\nonumber\\
  \label{eq:fullpsi2}
  &- \bar n(k)\bar n(k')\frac{\bra{\kvec,\kvec'}\hat V\psi
    \ket{\hvec,\hvec'}}{t(k) + t(k') -t(h)-t(h')}
\,,
\end{align}
where, in the simplest case, $t(k) = \hbar^2 k^2/2m$. In the
conventional Bethe-Goldstone equation, $\hat V$ was meant to be the
bare interaction operator $\hat v$. In FHNC-EL or parquet-theory,
$\hat v$ is supplemented by the induced interaction $\hat w_{\rm I}$ defined
in Eq. \eqref{eq:Scond}.  We can also have ``non-parquet'' diagrams --
in (F)HNC-EL these are due to ``elementary diagrams'' and
multiparticle correlations, while in the language of perturbation theory
these are particle-particle and particle-hole irreducible vertices.
Thus, in general, we may assume
\begin{equation}
  \hat V(i,j) = \hat v(i,j) + \hat w_{\rm I}(i,j) + \hat V_{\rm I}(i,j)\label{eq:Veff}
\end{equation}
where $\hat V_{\rm I}(i,j)$ is the set of irreducible diagrams. All three
sets are assumed to have the operator structure \eqref{eq:vop}.

The pair wave function $\psi$ is still a function of three momenta. On
the other hand, the variational wave function \eqref{eq:f_prodwave}
contains only functions that depend on the distance between two
particles.  To make a connection between perturbation theory and the
variational wave function, we must therefore approximate the pair wave
function by a quantity that depends only on the relative coordinate
(or momentum), \ie it has the feature
\[
  \bra{\kvec,\kvec'}\psi\ket{\hvec,\hvec'} = \frac{1}{N}\tilde\psi(q)
  \]
For local interactions, we then have also
\[
\bra{\kvec,\kvec'}v\psi\ket{\hvec,\hvec'} = \frac{1}{N}
\left[v(r)\psi(r)\right]^{\cal F}(q)\,.
\]
To have such a solution, the energy denominator coefficient
must be approximated by a function of momentum transfer
$q$. One option is to write Eq.~(\ref{eq:fullpsi2}) as
\begin{align}
   &\left[t(k) + t(k') -t(h)-t(h')\right]
\left[\bra{\kvec,\kvec'}\psi\ket{\hvec,\hvec'}-
\ovlp{\kvec,\kvec'}{\hvec,\hvec'}\right]\nonumber\\
  &=-\bar n(k)\bar n(k')\bra{\kvec,\kvec'}v\psi\ket{\hvec,\hvec'}
  \label{eq:BGpsi3}\,.
\end{align}
and then approximate the particle-hole energy differences by their
Fermi-sea average,
\begin{align}
  t(|\hvec+\qvec|) -t(h)&\approx 
\frac{\sum_{\hvec} \bar n(|\hvec+\qvec|) n(h) t(|\hvec+\qvec|) -t(h)}
     {\sum_{\hvec} \bar n(|\hvec+\qvec|) n(h)}\nonumber\\
     &= \frac{t(q)}{\SF(q)} \equiv \tF(q)\,.
\end{align}
This leads to
\begin{align}
  &\left[-\frac{\hbar^2}{m}\nabla^2 + v(r)\right]\psi(r)\nonumber\\
  &= -\frac{\rho}{\nu}
  \int d^3r' \ell^2(|\rvec-\rvec'|\KF)v(r')\psi(r')
  \label{eq:BGSchr}\,,
\end{align}
see Ref. \onlinecite{v3eos} for a different but equivalent way to
write this equation.  Eq.~(\ref{eq:BGSchr}) is very similar to the
Bethe-Goldstone equation for a pair of particles whose center of mass
momentum is zero. In that case, one obtains
\cite{BetheGoldstone57,FetterWalecka}
\begin{align}
  &\left[-\frac{\hbar^2}{m}\nabla^2 + v(r)\right]\psi(r)\nonumber\\
  &= -\frac{\rho}{\nu}
  \int d^3r' \ell(|\rvec-\rvec'|\KF)v(r')\psi(r')
  \label{eq:BGcmzero}\,.
\end{align}
The $G$-matrix is, in the local approximation, given by
\begin{equation}
  \hat G(\qvec) = \hat V(\qvec) - \int \frac{d^3q'}{(2\pi)^3}
         \hat V(|\qvec-\qvec'|)
         \frac{\hat G(\qvec')}{2\tF(q')}\,.\label{eq:BGlocal}
\end{equation}
The convolution product is best written in coordinate space,
\begin{equation}
\hat G(\rvec) = \hat V(\rvec) - \hat V(\rvec)\left[\frac{\hat G(\qvec)}{2\tF(q )}\right]^{\cal F}(\rvec)
\label{eq:BGrspace}
\end{equation}
where ${\cal F}$ stands for the Fourier transform \eqref{eq:ft} or
\eqref{eq:ft2}.  We have above not explicitly spelled out the operator
dependence which is implied. The equations are the same for
state-dependent interactions, they separate in the projector
representation $\{ \hat P_S, \hat P_{T+}, \hat P_{T-}\}$

We stress here that the local ``particle-hole'' interaction $\hat
V_{\rm p-h}(\qvec)$ entering the summation of ring ring diagrams must
{\em not\/} be identified with some local approximation of the
$G$-matrix. This is seen most easily in a self-bound system like
nuclear matter by the argument that the Fermi-sea average of the $G$
matrix should basically be the interaction correction to the binding
energy which is negative. On the other hand, the matrix element of the
central component of $\hat V_{\rm p-h}(\qvec)$ at the Fermi surface is
the interaction correction to the incompressibility which is positive
\cite{Shlomo06}. The problem is not as significant
for repulsive systems like neutron matter studied here or for
electrons \cite{LoB75}, we see, on the other hand, no reason to make
such unnecessary approximations.

The FHNC-EL equations lead to a slightly different form
\cite{fullbcs}, but note that FHNC sums more than just the
particle-particle ladders. We found, however, in our numerical
applications that the numerical solutions are very close. We shall,
therefore, not elaborate on this issue any further.  Diagrammatically
we can identify the pair wave function $\psi(r)$ with the direct
correlation function $\Gamma_{\!\rm dd}(r)$
\begin{equation}
  \psi(r) = \sqrt{1+\Gamma_{\!\rm dd}(r)}\,.\label{eq:psiGdd}
\end{equation}

\section{Beyond parquet: Including ``twisted chain'' diagrams}

\subsection{Low order analysis}
\label{sec:Goldstone}

We now turn to the main issue of this work, which is the diagrammatic
content and the treatment of the commutator terms introduced by the
need to symmetrize the operator product \eqref{eq:f_prodwave}.  For
this purpose, we utilize the correspondence between diagrams of the
cluster expansions for Jastrow-Feenberg wave functions and specific
approximations to Goldstone diagrams. Such a correspondence has been
observed a long time ago \cite{GGR,RIP79}. In a very vague language,
Jastrow-Feenberg diagrams and Goldstone diagrams can be identified by
absorbing the energy denominator in the interaction which then defines
a dimensionless function $h(r_{ij})$. What remains is only the
momentum flux and the Pauli operators.

The rules on how to translate a Goldstone
diagrams into a Jastrow-Feenberg diagram are then easily verified:
\begin{itemize}
\item{} Re-interpret each interaction line by a correlation
  function $h(r_{ij}) = f^2(r_{ij})-1$,
  \item{} Omit all energy denominators,
\item{} Each hole line turns into an exchange line $\ell(r_{ij}\KF)$
  where $\ell(x) = \frac{3}{x}j_1(x)$.
\item{} Each particle line turns into $\delta(r_{ij})-\ell(r_{ij}\KF)$
\end{itemize}
There cannot be an exact one-to-one correspondence because the wave
function \eqref{eq:Jastrow} or \eqref{eq:f_prodwave} is defined for
{\em any\/} correlation function whereas the sum of all Goldstone
diagrams converges towards the exact ground state. To make the
connection complete one must also include the optimization of the
correlations.

\begin{figure}[H]
 \includegraphics[width=0.3\columnwidth]{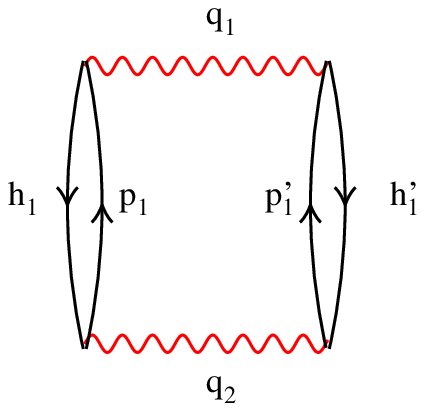}
 \hspace{0.05\columnwidth}\includegraphics[width=0.6\columnwidth]{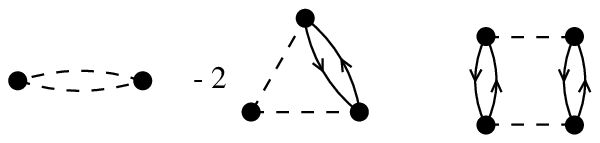}
 \caption{(color online) The left diagram is the second-order
   Goldstone diagram, the wiggly line represents an interaction. The
   three diagrams on the right hand side are the corresponding
   JF-diagrams, the usual diagrammatic conventions \cite{Johnreview}
   apply: The dashed line represent correlation factors $h(r_{ij})$
   and the oriented solid ones represent exchange lines
   $\ell(r_{ij}\KF)$.}
 \label{fig:v2fhnc}
 \end{figure}

To see how this works, we consider the simple second order
perturbation theory. To simplify the notation, we do this for central
interactions only. The left figure in
Fig. \ref{fig:v2fhnc} is the second order Goldstone diagram, the right
figure is what would result from the above operations. Note that the
first of the JF diagrams does not appear in cluster expansions of the
Jastrow-Feenberg wave function.

\begin{figure}[h]
  \centering
    \includegraphics[width=0.6\columnwidth]{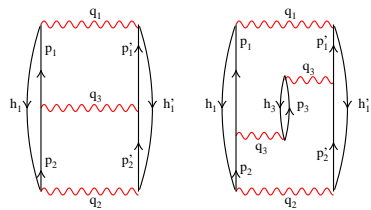}
    \caption{(color online) The simplest ladder diagrams. The left
      diagram is the ordinary 3-rung ladder, the right one contains an
      induced interaction. }\label{fig:v4twist}
\end{figure}

Let us next see how the localization procedure \eqref{eq:Scond} of
parquet theory works in the case of a simple ladder
diagram. Fig. \ref{fig:v4twist} shows the third order ladder diagram in
which the middle rung in the left diagram is replaced by an
induced interaction in the right diagram.

The exact form of the left diagram is, for local translationally
invariant interactions,
\begin{align}
  \frac{1}{N^3}\sum_{q_ih_1h_1'}&\tilde v(q_1)\frac{\bar n(\hvec_1+\qvec_1)\bar n(\hvec_1'-\qvec_1)}
       {e_{\hvec_1+\qvec_1}+e_{\hvec_1'-\qvec_1}-e_{h_1}-e_{h_1'}}
    \times\nonumber\\
    \times&\tilde v(\qvec_1-\qvec_2)
    \frac{\bar n(\hvec_1+\qvec_2)\bar n(\hvec_1'-\qvec_2')}
         {e_{\hvec_1+\qvec_2}+e_{\hvec_1'-\qvec_2}-e_{h_1}-e_{h_1'}}
   \tilde v(q_2)\,.
\end{align}
where $n(k) = \theta(\KF-k)$ is the Fermi distribution, and $\bar n(k)
= 1-n(k)$.  We can write the right diagram in Fig. \ref{fig:v4twist}
in a similar way
\begin{align}
  \frac{1}{N^3}\sum_{q_ih_1h_1'}&\tilde v(q_1)\frac{\bar n(\hvec_1+\qvec_1)\bar n(\hvec_1'-\qvec_1)}
       {e_{\hvec_1+\qvec_1}+e_{\hvec_1'-\qvec_1}-e_{h_1}-e_{h_1'}}
    \times\nonumber\\
    \times&\tilde w(\qvec_1-\qvec_2,e_{\hvec_1+\qvec_1}+e_{\hvec_1'-\qvec_2}-e_{h_1}-e_{h_1'})\times\nonumber\\
    \times&\frac{\bar n(\hvec_1+\qvec_2)\bar n(\hvec_1'-\qvec_2')}
         {e_{\hvec_1+\qvec_2}+e_{\hvec_1'-\qvec_2}-e_{h_1}-e_{h_1'}}
   \tilde v(q_2)\,.
\end{align}
with an energy-dependent interaction
\begin{equation}
  \tilde w(q,\hbar\omega) = -\tilde v^2(q)
  \frac{1}{N}\sum_h\frac{1}{e_{\hvec+\qvec}-e_{h}+\hbar\omega}\,.
  \label{eq:wofqw}
\end{equation}
The localization procedure of parquet theory replaces the
energy-dependent induced interaction $\tilde w(q,\hbar\omega)$ by an
energy-independent interaction which is constructed from $\tilde
w(q,\hbar\omega)$ by evaluating this quantity at an averaged frequency
$\bar\omega(q)$, \ie
\begin{equation}
  \tilde w(q) =
  \tilde w(q,\hbar\bar\omega(q))\,,\label{eq:wofq}
\end{equation}
see Eq. \eqref{eq:Scond}. Once $\tilde w(q,\hbar\omega)$ has been
replaced by $\tilde w(q) = \tilde w(q,\hbar\bar\omega(q))$, the
combination $\tilde v(q) + \tilde w(q)$ can be used as an effective
interaction in the Bethe-Goldstone equation, this is exactly the
connection between the parquet-diagram and the (F)HNC-EL view of
ground state correlations.

\begin{figure}[h]
  \centering
    \includegraphics[width=0.9\columnwidth]{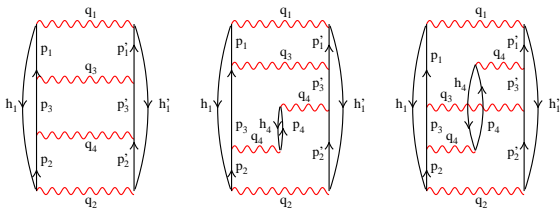}
    \caption{(color online) Fourth-order ladder diagrams, including a
      ``twisted chain'' diagram. The left diagram is the 4-body
      ladder, in the middle diagram one of the interactions is
      replaced by an induced interaction, and the right one the
      twisted version.}\label{fig:v5twist}
\end{figure}

In the next order, shown in Fig. \ref{fig:v5twist}, we first see the
possibility of ``twisting'' chain diagrams.

The left and the middle diagram can again be combined by introducing
the energy-dependent induced interaction $\tilde w(q,\hbar\omega)$ which is
then, in the local approximation, replaced by $\tilde w(q)$ as in Eq. 
\eqref{eq:wofq}.

The third diagram, although it has the same components, is by its very
definition not a parquet diagrams, and cannot be represented in terms
of the energy-dependent interaction $\tilde w(q,\hbar\omega)$. Apply
now the rules, discussed in connection with Fig. \ref{fig:v2fhnc}, for
how to identify Goldstone and Jastrow-Feenberg diagrams and
re-interpret the second and third diagram shown in
Fig. \ref{fig:v5twist} as Jastrow-Feenberg diagrams.  We then find
that these two diagrams have indeed the same value, \ie the
Jastrow-Feenberg wave function suggests to approximate these two terms
by the same quantity. We can therefore conclude that, {\em if\/} we
want to approximate the cross-going portion in the third diagram by a
static interaction, this should be the same as the induced interaction
$\tilde w(q,\omega)$. Moreover, the equivalence of Jastrow-Feenberg
theory and the local parquet theory shows that the FHNC-EL
approximation contains {\em both\/} terms.

The above analysis is valid only for state-independent interactions or
correlations or, in other words, for the configuration-space
components of this diagram. There is no reason that the same argument
should work for the spin/isospin components. Assume therefore now that
interactions are state dependent. For the present purpose, it is best
to represent them in the $\{\1,\hat L, \hat T\}$ basis, then the
operators on the interactions with momentum transfer $q_4$ must be the
same, say $\hat O_\alpha$ and the operator associated with the
interaction with momentum transfer $q_4$ be $\hat O_{\beta}$. We were
above led to the conclusion that variational wave functions suggest
the approximation that the coordinate or momentum dependence of the
sub-diagrams with momentum transfer $q_4$ are the same, but the
operator dependence has to be included more carefully.  The second
diagram in Fig. \ref{fig:v5twist} has then the operator structure
\[\Tr_4 \left[\hat O_\alpha(14)\hat O_\alpha(42)\hat O_\beta(12)\right]\]
whereas the third diagram has the operator order
\[\Tr_4 \left[\hat O_\alpha(14)\hat O_\beta(12)\hat O_\alpha(42)\right]\,,\]
\ie the combination of these two operators is exactly the symmetrized
product.

The way to correct the parquet-equations \cite{SmithSpin} or the
version of the (F)HNC equations that ignores all commutators
\cite{FantoniSpins} is therefore to add the commutator of these two
terms.
\[\frac{1}{2}\Tr_4 \left[\hat O_\alpha(14)\left[\hat O_\beta(12),
    \hat O_\alpha(42)\right]\right]\,.\]

\subsection{Twisted chains corrections to the Bethe Goldstone equation}
\label{ssec:twist}

We now turn to including the ``twisted chain'' diagrams in the
localized Bethe-Goldstone equation.

We have two diagrammatic elements: The bare interaction $\hat
v(\qvec)$ which is completely irreducible and the induced interaction
$\hat w_{\rm I}(\qvec)$, which is particle-hole reducible and comes from the
FHNC-EL equations or is constructed by means of the ``average energy''
approximation \eqref{eq:wofq}. The bare interaction comes always in
combination with the induced interaction, we will depict the sum of
these two as a magenta wiggled line.  For the following calculations,
we assume that both the bare interaction $\hat v(i,j)$ and the induced
interaction $\hat w(i,j)$ are given in the operator basis $\{\1, \hat
L, \hat T\}$. In that basis, it is sufficient to consider chains of
two elements as shown in Fig. \ref{fig:v5twist}, the longer chains
left and right of the particle-hole bubble $\{p_4,h_4\}$ can be summed
into one term.

The ``cross-going'' diagrams, \ie those of the topology of the third
diagram shown in Fig. \ref{fig:v5twist} must all be particle-hole
reducible, we will depict these as blue wiggled lines.  As we have
shown above, the Jastrow-Feenberg wave function suggests that
configuration space of the second and the third diagram in
Fig. \ref{fig:v5twist} are the same and all we need to do
is to include the commutators.

Let us assume that the operator connected with the interaction line
$\hat v + \hat w_{\rm I}$ in Fig. \ref{fig:v5twist} is $\hat O_v(i,j)$. The
operator connected with the induced interaction is $\hat
O_w(i,j)$. These operators are either $\hat L(i,j) =
\sigma_\alpha(i)t_{\alpha\beta}^{(L)}(\hat\rvec)\sigma_\beta(j)$,
$t_{\alpha\beta}^{(L)}=\hat r_\alpha\hat r_\beta $ or $\hat T(i,j)
=\sigma_\alpha(i)t_{\alpha\beta}^{(T)}(\hat\rvec)\sigma_\beta(j)$,
$t_{\alpha\beta}^{(T)}=\hat r_\alpha\hat r_\beta -
\delta_{\alpha\beta} $.  We will also need the same set of operators in
momentum space, the unit vector $\hat\rvec$ is then replaced by
$\hat\qvec$.  We label the {\em external\/} points with
$a$, $b$ and the {\em internal\/} points with numbers.  The {\em
  correction\/} to the unsymmetrized operator product is then given by
the commutator
\begin{equation}
  \frac{1}{2}\Tr_{1}\left[\hat O_w(a,1)\left[\hat O_v( a,b),\hat O_w(1,b)\right]
    \right]\label{eq:twistvw}
\end{equation}
where the $\hat O_w$ are the spin-operators associated with the induced
interaction $\hat w_{\rm I}$ and $\hat O_v$ are those associated with $\hat
v + \hat w_{\rm I}$.  The commutator with the central operator is evidently
zero. In what follows, we will also need the relationships

\begin{subequations}
\begin{eqnarray}
  \hat L^2 &=& \hat P_S + \hat P_{T+}+\hat P_{T-} = \1\,,\label{eq:LL}\\
  \hat T^2 &=& 4\hat P_S + 4\hat P_{T-} = 2\1-2\hat L\,,\label{eq:TT}\\
  \hat L\hat T  &=& 2\hat P_S -2\hat P_{T-} = -\hat T\,.\label{eq:LT}  
\end{eqnarray}
\end{subequations}

For both the longitudinal and the transverse operators, we have
  $ \sum_\beta t_{\alpha\beta}t_{\beta\mu} = t_{\alpha\mu}$, we get therefore
for (\ref{eq:twistvw})
\begin{align}
    &\Tr_{1}\left[\sigma_\alpha(a)t_{\alpha\beta}^{(w)}\sigma_\beta(1)
      \sigma_\gamma(a)t_{\gamma\delta}^{(v)}\sigma_\delta(b)
      \sigma_\lambda(1)t_{\lambda\mu}^{(w)}\sigma_\mu(b)\right]\nonumber\\
  &- \nu O_v(a,b)O_w(a,b)
    \nonumber\\
    &=-2\nu O_v(a,b)O_w(a,b) + 2 \nu t_{\alpha\beta}^{(w)}t_{\alpha\beta}^{(v)}
    \,.\nonumber 
  \end{align}

We must now distinguish three cases:
\begin{itemize}
\item{} Both $\hat O_v(a,b) = \hat O_w(a,b) = \hat L(a,b)$. Use Eq. (\ref{eq:LL})
  \begin{equation}
    -2\nu \hat L^2(a,b) + 2 \nu t_{\alpha\beta}^{(L)}t_{\alpha\beta}^{(L)} = 0,.
    \label{eq:LLtwist}
  \end{equation}
\item{} $\hat O_v(a,b) = \hat L(a,b)$ and $\hat O_w(a,b) = \hat T(a,b)$.
  Use Eq. (\ref{eq:LT})
  \begin{equation}
    -2\nu \hat L(a,b)\hat T(a,b) + 2 \nu t_{\alpha\beta}^{(L)}t_{\alpha\beta}^{(T)} = 2\nu\hat T(1,2)\,.
    \label{eq:LTtwist}
  \end{equation}

\item{} Both $\hat O_v(a,b) = \hat O_w(a,b) = \hat T(a,b)$. Use
  Eq. (\ref{eq:TT})
  \begin{equation}
      -2\nu \hat T(a,b)^2 + 2 \nu t_{\alpha\beta}^{(T)}t_{\alpha\beta}^{(T)} = 4\nu\hat L(a,b),.\label{eq:TTtwist}
  \end{equation}
  \end{itemize}

\begin{figure}[h]
  \centering
    \includegraphics[width=0.9\columnwidth]{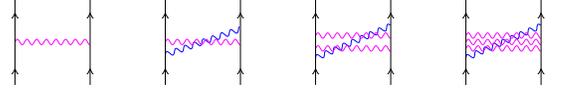}
    \caption{\small Examples where the reducible induced interaction
      crosses more than one rung. The magenta line represents the sum
      $\hat v + \hat w_{\rm I}$. The rungs can all be summed to the
      $G$-matrix.  }  \label{fig:vn_twist}
 \end{figure}

So far we have only considered the simplest process. Next, consider
the series shown in Fig. \ref{fig:vn_twist}. The summation of these
diagrams is necessary to deal with short-ranged correlations.

The diagram with $n$ rungs and one crossing has the spin-operator
structure
\begin{eqnarray}
  &&\Tr_{1} \biggl[\hat O_w(a,1)\hat O_{v_1}(a,b)
    \dots,\hat O_{v_n}(a,b)\hat O_w(b,1)\nonumber\\
    &&-\hat O_{v_1}(a,b)\dots,\hat O_{v_n}(a,b)\hat O_w(a,1)
  \hat O_{w}(b,1)\biggr]\nonumber
\end{eqnarray}
where the $O_{v_i}$ are the spin-operators connected to the $n$ rungs.
These are {\em a priori\/} from the set $\{\1, \hat L, \hat T\}$.  We
rewrite the product $\hat O_{v_1}(a,b)\dots,\hat O_{v_n}(a,b)$ in
terms of the projection operators \eqref{eq:projectors}. All of these
operators have, in coordinate space, the same spatial argument
$\hat\rvec$. They are therefore idempotent and, hence, the product
$\hat O_{v_1}(a,b)\dots,\hat O_{v_n}(a,b)$ can be rewritten as a
linear combination of the projection operators \eqref{eq:projectors}
which, at the end, is transformed to a linear combination of the set
$\{\1, \hat L, \hat T\}$. The conclusion is that sum of all magenta
lines in Fig. \ref{fig:vn_twist} can be represented by the $G$ matrix.

\begin{figure}[h]
  \centering
    \includegraphics[width=0.9\columnwidth]{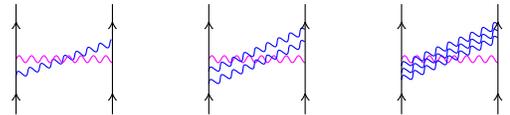}
    \caption{\small The summation of more than one cross-going lines.
 }     \label{fig:wn_twist}
 \end{figure}

Similarly, we can calculate the set of diagrams shown in
Fig. \ref{fig:wn_twist}.
Note that we can here, according to the above, interpret the magenta wavy
line as a component of the $G$-matrix.
Then the operator form of a diagram with $n$ crossing rungs in
Fig. \ref{fig:wn_twist} is
\begin{widetext}
\begin{eqnarray}
  &&\Tr_{1\dots n}\biggl[\hat O_{w_1}(a,1)\dots\hat O_{w_n}(a,n)
    \hat O_v(a,b)\hat O_{w_1}(1,b)\dots\hat O_{w_n}(n,b)\nonumber\\
    &&\qquad-\hat O_v(a,b) \hat O_{w_1}(a,1)\hat O_{w_1}(1,b)\dots
    \hat O_{w_n}(a,n)\hat O_{w_n}(n,b)\biggr]\label{eq:multitwist}\,.
\end{eqnarray}
To evaluate this expression, we use
\begin{equation}
  \hat O_v(a,b)=\frac{1}{\nu}
  \Tr_{n+1}\biggl[\hat O_v(a,n+1)\hat O_v(n+1,b)\biggr]\,.
\end{equation}
Therefore, Eq. (\ref{eq:multitwist}) can be rewritten in the form
\begin{eqnarray}
  &&\frac{1}{\nu}\Tr_{1\dots n+1}\left[\hat O_v(n+1,b)\hat O_{w_1}(a,1)
    \hat O_{w_1}(1,b)
    \dots\hat O_{w_n}(a,n)\hat O_{w_n}(1,n)\hat O_v(a,n+1)\right]\nonumber\\
    &&\qquad-\Tr_{1\dots n}\left[\hat O_v(a,b) \hat O_{w_1}(a,1)\hat O_{w_1}(1,b)\dots
    \hat O_{w_n}(a,n)\hat O_{w_n}(n,b)\right]\nonumber\\
  &=&\frac{\nu^n}{\nu}\Tr_{n+1}\left[\hat O_v(n+1,b)\hat O_{w_1}(a,b)
 \dots\hat O_{w_n}(a,b)\hat O_v(a,n+1)\right]\nonumber\\
    &&\qquad-\nu^n\left[\hat O_v(a,b) \hat O_{w_1}(a,b)\dots O_{w_n}(a,b)\right]
\end{eqnarray}
\end{widetext}
We can now use the same argument as above to show that the product
$\hat O_{w_1}(a,b)\dots O_{w_n}(a,b)$ can be written as a linear
combination of the operator set $\{\1, \hat L, \hat T\}$.  Summing
over all the blue lines in Fig. \ref{fig:wn_twist} gives just another
Bethe-Goldstone equation in which the $\hat v+\hat w$ is replaced by
the $\hat w$. This defines a modified $G$-matrix, say $\hat G^{(w)}$,
where all the rungs are just induced interactions.

Summarizing, the Bethe-Goldstone equation \eqref{eq:BGlocal} with the
effective interaction \eqref{eq:Veff} is supplemented by a second
equation that sums the rungs containing only induced interaction lines
\begin{align}
  \hat G^{(w)}(q) &= \hat w_{\rm I}(q) - \int \frac{d^3q'}{(2\pi)^3}
         \hat w_{\rm I}(\qvec-\qvec')
         \frac{\hat G^{(w)}(q')}{2\tF(q')}\,.\label{eq:BGP}
\end{align}

Along with the calculation of the $G$-matrix we obtain the pair wave
function $\tilde \psi(q)$, Eq. \eqref{eq:BGSchr} and an analogous
quantity $\tilde \psi^{(w)}(q)$ corresponding to $\tilde G^{(w)}(q)$.
  
The above integral equations \eqref{eq:BGlocal} and \eqref{eq:BGP} sum
the two types of ladder diagrams.  The purpose of the summations shown
in Figs. \ref{fig:vn_twist} and \ref{fig:wn_twist} was to demonstrate
that we can replace the sum of all magenta wavy line by $\hat
G(\rvec)$ and the sum of all blue lines by $\hat G^{(w)}(\rvec)$.

\subsection{The irreducible part of the interactions}
\label{ssec:derivations}

Eqs.  \eqref{eq:BGlocal} and \eqref{eq:BGP} are solved in the $\{P_S,
P_{T+}, P_{T-}\}$ basis, we obtain therefore the operators in the
representation
\begin{align}
  \hat G(\rvec) &= G_S(r)\hat P_S + G_{T+}(r)\hat P_{T+}+ G_{T-}(r)\hat P_{T-}
  \nonumber\\
  &= \sum_\alpha G_\alpha(r)\hat P_\alpha\label{eq:Gproj}
\end{align}
and the same form for $\hat G^{(w)}(\rvec)$.  To calculate the
correction $\hat V_{\rm I}(\qvec)$ we now go back to the analysis of
Fig. \ref{fig:v5twist}. We first rewrite both quantities in the basis
$\{\1, \hat L, \hat T\}$.  We can then use the coupling coefficients
derived in Eqs. (\ref{eq:LLtwist}) - (\ref{eq:TTtwist}).  We can then
write
\begin{widetext}
\begin{equation}
  \hat V_{\rm I}(\qvec)=-\sum_{\alpha,\beta}
  \int \frac{d^3q'}{2(2\pi)^3\rho}\tilde G_\alpha(|\qvec-\qvec'|)
  \frac{\tilde G^{(w)}_\beta(q')}{2\tF(q')}\Tr_1
  \left[\hat O_\beta(a,1)\left[\hat O_\alpha(a,b),
        \hat O_\beta(1,b)\right]\right]
    \end{equation}
\end{widetext}
where it is implied that the operators $O_\alpha(a,b)$ are from the
set $\{\1, \hat L, \hat T\}$. Of course, the commutator with the
central operator $\1$ is zero. Using
Eqs. \eqref{eq:LLtwist}-\eqref{eq:TTtwist} gives $\hat V_{\rm I}(\rvec)$ in
the same basis, we must therefore transform back to $\{P_S, P_{T+},
P_{T-}\}$ basis.  From Eqs.  \eqref{eq:Gproj} we finally obtain $\hat
G(\rvec)$ and $\hat G^{(w)}(\rvec)$ in the projector basis $\{P_S,
P_{T+}, P_{T-}\}$.
\begingroup 
\allowdisplaybreaks
\begin{subequations}
  \begin{align}
    V_{\rm I}^{(S)}(r) =&-\frac{1}{8}G_S(r)(-3\psi^{(w)}_{S}(r)+2\psi^{(w)}_{T+}(r)+\psi^{(w)}_{T-}(r))
    \nonumber\\
    &-\frac{1}{4}G_{T+}(r)(\psi^{(w)}_{S}(r)-\psi^{(w)}_{T-}(r))\label{eq:GGPs}\\
    &-\frac{1}{8}G_{T-}(r)(\psi^{(w)}_{S}(r) -2 \psi^{(w)}_{T+}(r)+\psi^{(w)}_{T-}(r))\,,\nonumber\\
    V_{\rm I}^{(T+)}(r)=&-\frac{1}{8}G_{S}(r)(\psi^{(w)}_{S}(r)-\psi^{(w)}_{T-}(r))\nonumber\\
    &+\frac{1}{8}G_{T-}(r)(\psi^{(w)}_{S}(r)-\psi^{(w)}_{T-}(r))\label{eq:GGPp}\,,\\
    V_{\rm I}^{(T-)}(r)=&-\frac{1}{8}G_S(r)(\psi^{(w)}_{s} - 2\psi^{(w)}_{T+}(r)+\psi^{(w)}_{T-}(r))
    \nonumber\\
    &+\frac{1}{4}G_{T+}(r)(\psi^{(w)}_{S}(r)-\psi^{(w)}_{T-}(r))\label{eq:GGPm}\\
    &-\frac{1}{8}G_{T-}(r)(\psi^{(w)}_{S}(r) + 2\psi^{(w)}_{T+}(r)-3\psi^{(w)}_{T-}(r))
    \nonumber\,.
  \end{align}
\end{subequations}
\endgroup
Eqs. \eqref{eq:GGPs}-\eqref{eq:GGPm} show exactly the conclusion drawn
from the analysis of the symmetrized operator product wave function:
The process described by diagrams of the kind discussed in
Figs. \ref{fig:vn_twist} and \ref{fig:wn_twist} mix interaction
components in different channels.  Self-consistency is obtained by
inserting the irreducuble terms $V_{\rm I}^{(\alpha)}(r)$ in the effective
interaction \eqref{eq:Veff} and repeating the process to convergence.

\section{Results}

\label{sec:results}

We have chosen in this work to study neutron matter for a number of
reasons. Neutron matter is, apart from the complications arising from
the state-dependence of the interactions, one of the simplest systems
of interest. As opposed to liquid $^3$He and nuclear matter, neutron
matter is not self-bound. A self-bound Fermi system has necessarily at
least two spinodal points below saturation density. An immediate
consequence of that is that the equation of state is a non-analytic
function of the density. Therefore, any expansion of the equation of
state in powers of the density cannot converge up to equilibrium
density. This complication does not exist in neutron matter and we can
focus on the problem at hand, which is the treatment of
operator-dependent correlations.

A consequence of the simplicity of neutron natter is, of course, that
relatively primitive approximations can lead, for some quantities, to
reasonable results. This is particularly true for the equation of
state because the error in the energy is of second order in the error
in the wave function. We must therefore look at quantities that depend
sensitively on the quality of the treatment of the many-body
problem. These are mostly effective interactions which are the
essential input for studying pairing properties (See
Refs. \onlinecite{SchuckBCS2018,SedrakianClarkBCSReview} or
\onlinecite{ECTSI} for review articles) and the density response of
neutron matter which have been discussed for decades
\cite{Wam93,Benhar2009,Lovato2013}.

We have carried out calculations for the Reid $v_6$ interaction
\cite{Day81} and the $v_6$ version of the Argonne interaction
\cite{AV18}.  The results are very similar and no insight is gained
from comparing these two interactions. We therefore report results for
the Reid potential only in the density regome $0.25\,{\rm fm}^{-1}\le
\KF\le 1.8\,{\rm fm}^{-1}$. The calculations to be presented here
refer to what we called in Ref. \onlinecite{v3eos} the ``parquet//1''
version.  The approximation goes beyond Jastrow-Feenberg in the sense
that propagator corrections are included in both the particle-particle
and the particle-hole channels. The notion ``//1'' refers to the
inclusion of first-order exchange diagrams.  These are necessary to
have a reasonably good agreement with the long-wavelength limit $\hat
V_{\rm p-h}(0+)$ and the Fermi-Liquid parameters from hydrodynamic
derivatives, see Refs. \onlinecite{fullbcs} and \onlinecite{v3eos} for
a discussion. Our calculations to be presented here go beyond the work
of Ref. \onlinecite{v3eos} by including diagrams that would correspond
to non-parquet diagrams in the language of perturbation theory, or to
commutator diagrams in the language of the variational
Jastrow-Feenberg method.

\subsection{Interaction corrections}

\begin{figure}[H]
   \centerline{\includegraphics[width=0.7\columnwidth,angle=270]{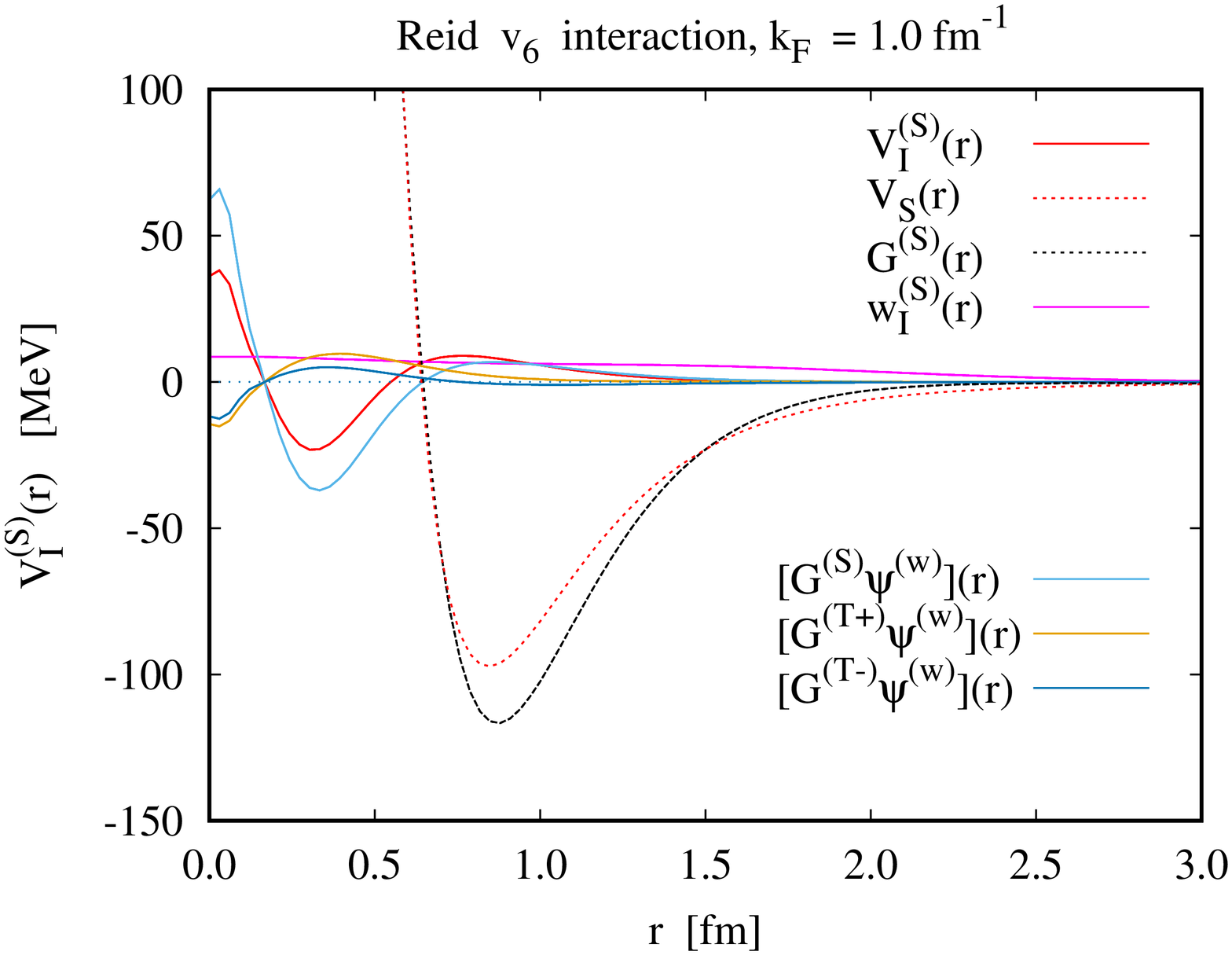}}
   \centerline{\includegraphics[width=0.7\columnwidth,angle=270]{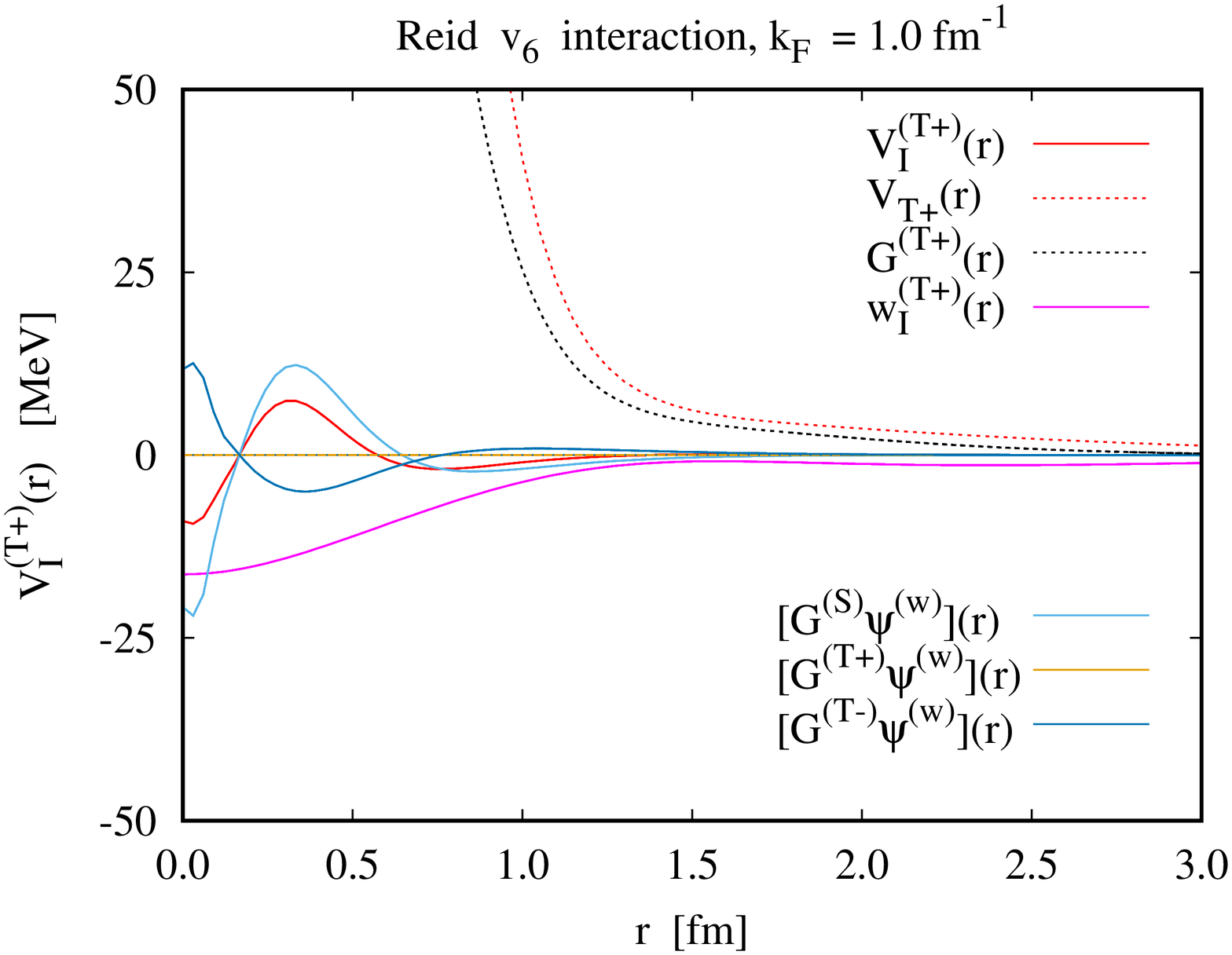}}
   \centerline{\includegraphics[width=0.7\columnwidth,angle=270]{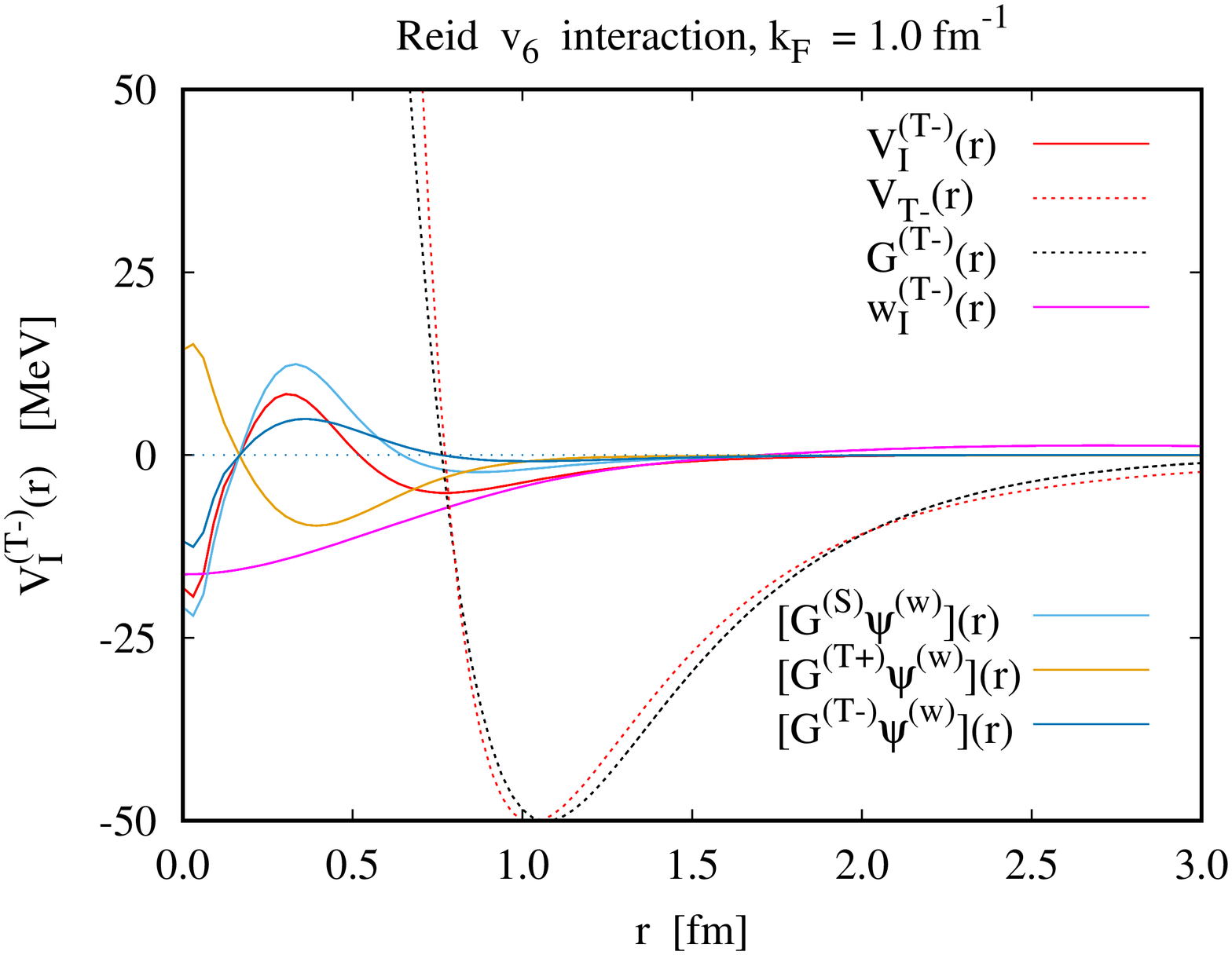}}
   \caption{(color online) The figures show, for the Reid $v_6$
     interaction at $\KF = 1.0\,\mathrm{fm}^{-1}$ the ``twisted
     chain'' correction $V_{\rm I}(r)$ (red) to the effective interactions,
     the induced interaction $w_{\rm I}(\rvec)$ (dark blue), the $G$-matrix
     (black dashed) in the local approximation (\ref{eq:BGlocal}) and the
     individual components of $V_{\rm I}(r)$ (light blue, beige, and dark blue lines) in the projector channels $S$, $T+$, and $T-$. We
     also show for reference the bare interaction in the same channels
     (red dashed lines).}
\label{fig:Vtwist10}
\end{figure}

One expects the most pronounced consequence of including ``twisted
chain'' diagrams in coordinate space at short and intermediate
distances.  Figs. \ref{fig:Vtwist10} show, at $\KF =
1.0\,\mathrm{fm}^{-1}$, the $G$-matrix in the local approximation
(\ref{eq:BGlocal}), the induced interaction $w_{\rm I}(r)$, and the
``twisted chain'' correction $V_{\rm I}(r)$. We also show the individual
components that were spelled out in
Eqs. \eqref{eq:GGPs}-\eqref{eq:GGPm}. For example
$\left[G^{(S)}\psi^{(w)}\right](r)$ shows the contribution from the
first line in Eq. \eqref{eq:GGPs},
$\left[G^{(T+)}\psi^{(w)}\right](r)$ the one from the second line and
$\left[G^{(T-)}\psi^{(w)}\right](r)$ the last term. The corresponding
information for the $T+$ and the $T-$ projections is shown in the
second and third figure, note that
$\left[G^{(T+)}\psi^{(w)}\right](r)$ has no component in the $T+$
channel.

In all three channels we observe the same features: the induced
interaction $w_{\rm I}(r)$ is rather smooth and relatively long-ranged
whereas the non-parquet diagram contributions are localized at short
and intermediate distances; this is similar to the contribution from
``elementary diagram'' and three-body correlations in quantum fluids.
The reason for this is simply the fact that $V_{\rm I}(r)$ falls off roughly
like the product of the interaction and $\psi^{(w)}(r)$.

In the singlet channel, the non-parquet corrections practically double
the repulsive induced interaction around the potential minimum, it
appears that this is a direct consequence of the large hard core of
the triplet channel potentials that is mixed into the singlet
channel. What is more important is that $V_{\rm I}(r)$ is in all three
channels comparable to the induced interaction. On the other hand, the
effect is practically irrelevant in the triplet channels because all
many-body corrections are overwhelmed by the larger core size of the
bare interaction.

\begin{figure}[H]
   \centerline{\includegraphics[width=0.7\columnwidth,angle=270]{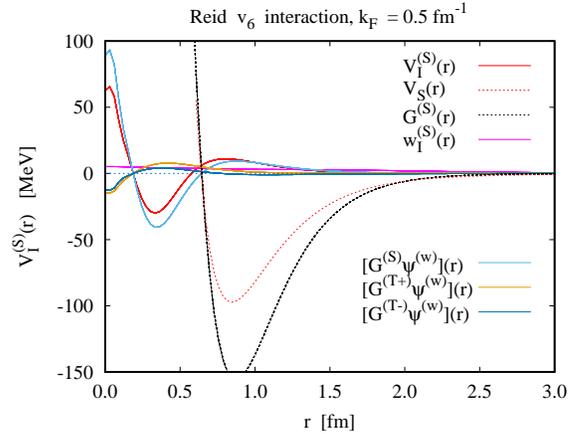}}
   \caption{(color online) Same as Fig \ref{fig:Vtwist10} for $\KF =
     0.5\,\mathrm{fm}^{-1}$. Only the singlet channel is shown.}
\label{fig:Vtwist05}
  \end{figure}

The situation changes rather drastically at lower densities. We show
in Fig. \ref{fig:Vtwist05} the individual components of the
interaction for $\KF = 0.5\,\mathrm{fm}^{-1}$.  In
the singlet channel, the $V_{\rm I}(r)$ is much larger than the
induced interaction $w_{\rm I}(r)$ and is, therefore, the dominant many-body
effect. The $G$-matrix becomes significantly more attractive in the
spin-singlet channel. The reason for this is found in the fact that
the bare $S$-wave interaction is, with a scattering length of $a_0
\approx -18.7\ $fm \cite{PhysRevLett.83.3788}, rather attractive and
close to a bound state. As a consequence, the pair wave function
$\psi(r)$ can change substantially if the interaction is only slightly
changed, this is the reason for the rather large nearest-neighbor peak
seen in Fig.  \ref{fig:Reid}. The large nearest neighbor peak has, in
turn, the consequence that the $G$-matrix becomes significantly more
attractive than the bare interaction. Of course, many-body effects and
the Pauli-principle still play the dominant role in determining the
pair wave function: The zero-energy $S$-wave scattering function has a
nearest neighbor peak of about 12, it is therefore nowhere close to
the in-medium pair wave function.

On the other hand, the correction from both the induced interaction
$\hat w_{\rm I} (r)$ and the ``twisted chain'' diagrams in the triplet
channels is again overwhelmed by the the large core size of the bare
interaction and therefore not shown.
 
\subsection{Correlation functions}

\begin{figure}[H]
   \centerline{\includegraphics[width=0.7\columnwidth,angle=270]{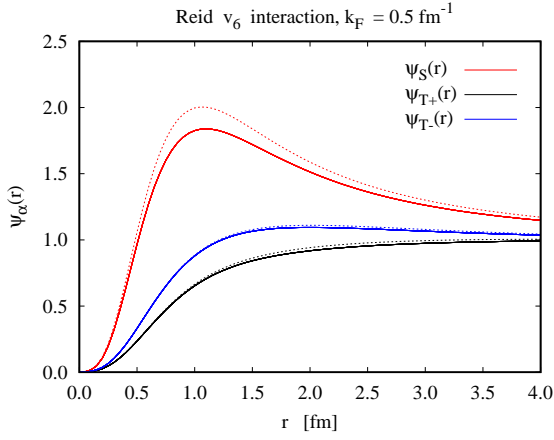}}
   \caption{(color online) The figures show, for the Reid $v_6$
     interaction at $\KF = 0.5\,\mathrm{fm}^{-1}$, the pair wave
     function $\psi_\alpha(r)$ in the three projector channels
     $S$ (red), $T+$ (black), and $T-$ (blue). The dashed lines
     show the parquet//1 case. Each corresponding solid line
     represents what is obtained if non-parquet diagrams are included.
   }
\label{fig:psitwist05}
  \end{figure}

To document the sensitivity of the pair correlations to the treatment
of many-body correlations, we show in Fig. \ref{fig:psitwist05} the
pair wave function $\psi_\alpha(r)$ in the three channels $\{S, T+,
T-\}$.  By adding the non-parquet contributions to the irreducuible
interaction, the peak in the pair wave function is reduced by about 10
percent. The effect is easily understood by the fact that the
irreducible diagrams mix a part of the more repulsive spin-triplet
interactions into the spin-singlet channel. The change is visible but
much more moderate in the spin-triplet channels which is consistent
with our findings on the effective interactions in
Figs. \ref{fig:Vtwist10} and \ref{fig:Vtwist05}. The effect becomes
larger at low densities because the attractive induced interaction
becomes weaker whereas the repulsive non-parquet corrections remain
roughly the same.

The strongly attractive $S$-wave interaction has led to discussions of
a potential BCS-BEC crossover in low-density neutron matter
\cite{NuclBCSBEC,SchuckBCS2018}, our findings would suggest that
many-body effects can somewhat reduce such a crossover. It
must be kept in mind, however, that the repulsive interaction in the
spin-triplet channels must not be neglected; it is responsible for
stabilizing neutron matter. A model system of nucleons interacting in
all channels with the $S$ wave interaction would have a very low
density spinodal point and would be unstable at any density that might
be of interest for the structure of neutron stars.

\subsection{Effective interactions}

\begin{figure}[H]
    \centerline{\includegraphics[width=0.65\columnwidth,angle=270]%
      {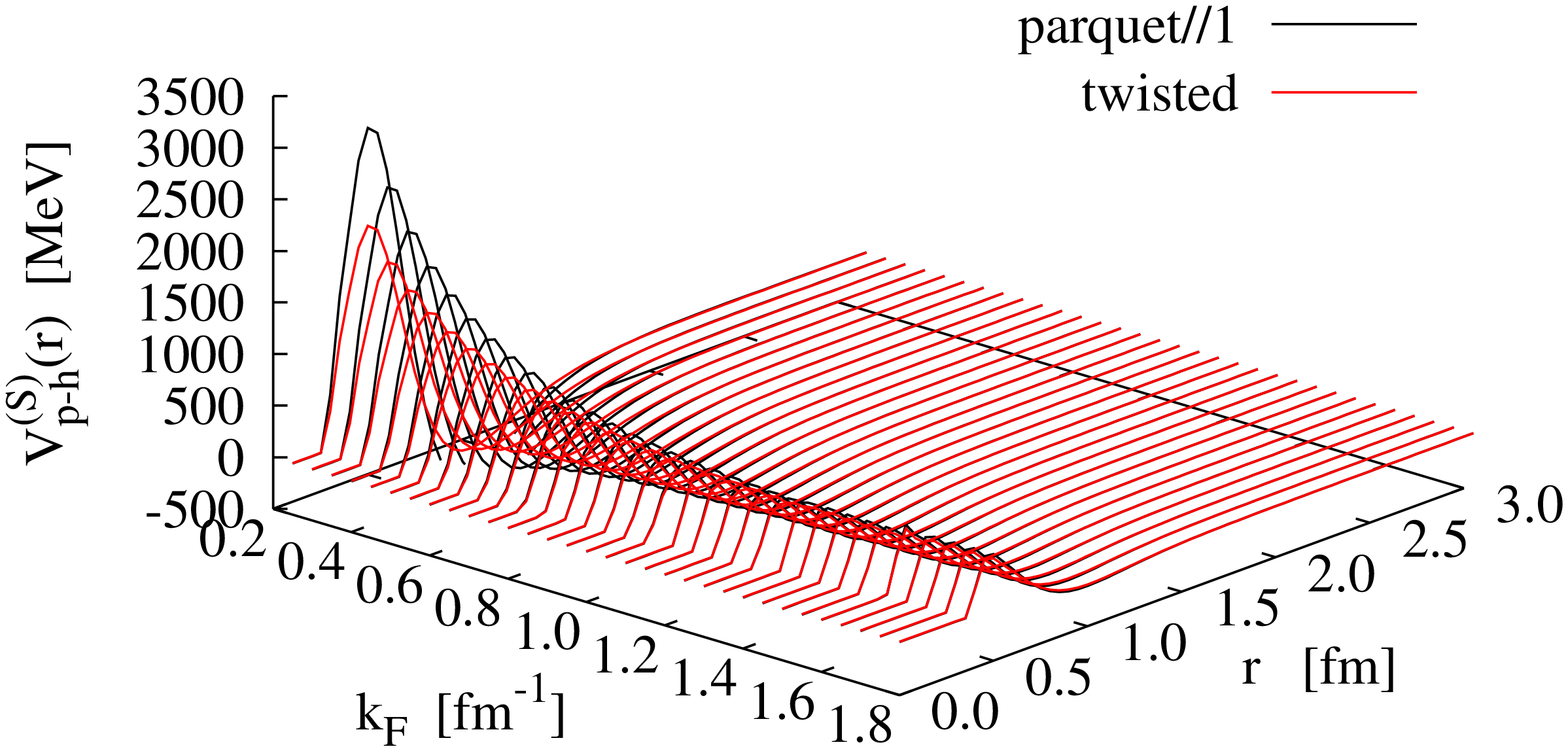}}
    \centerline{\includegraphics[width=0.65\columnwidth,angle=270]%
      {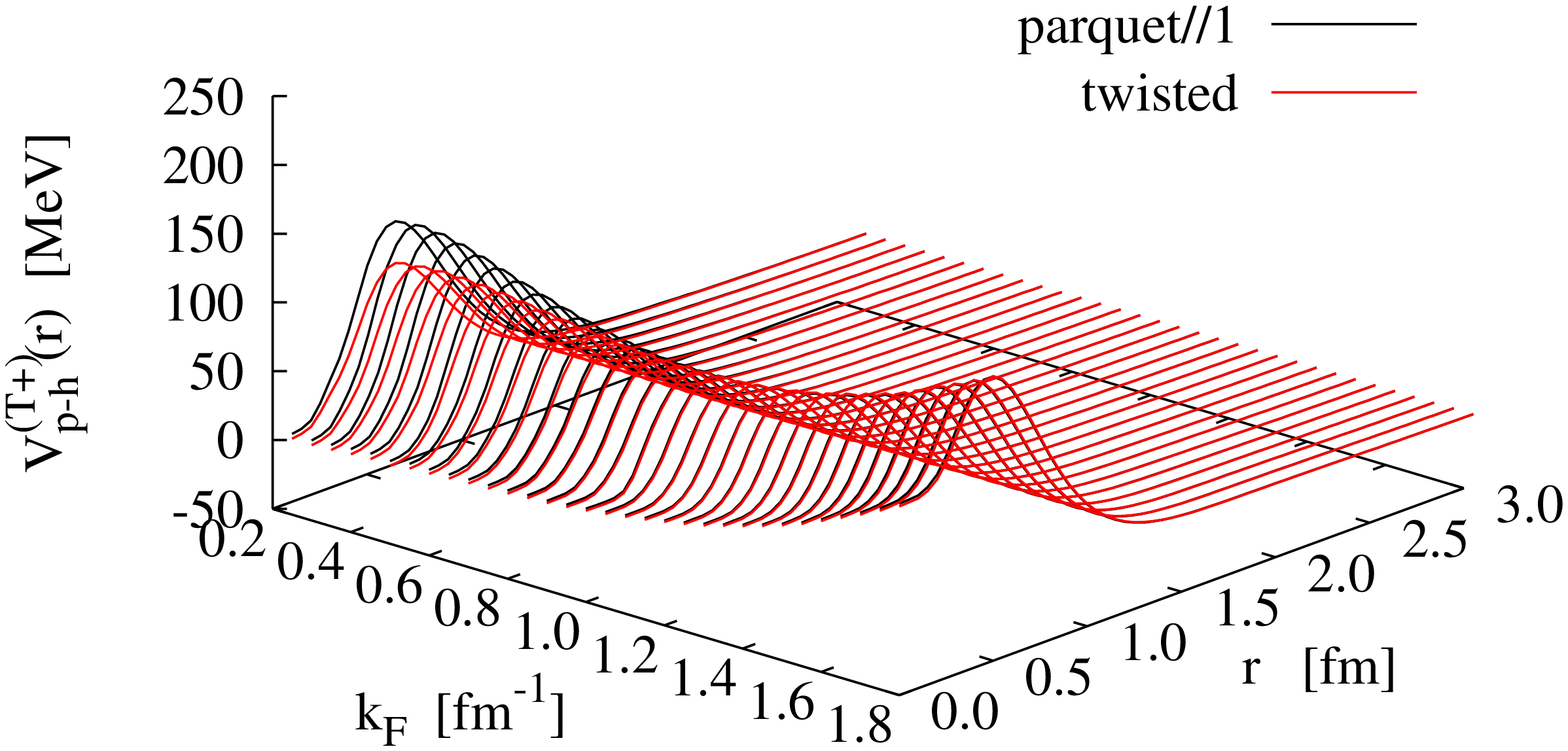}}
    \centerline{\includegraphics[width=0.65\columnwidth,angle=270]%
      {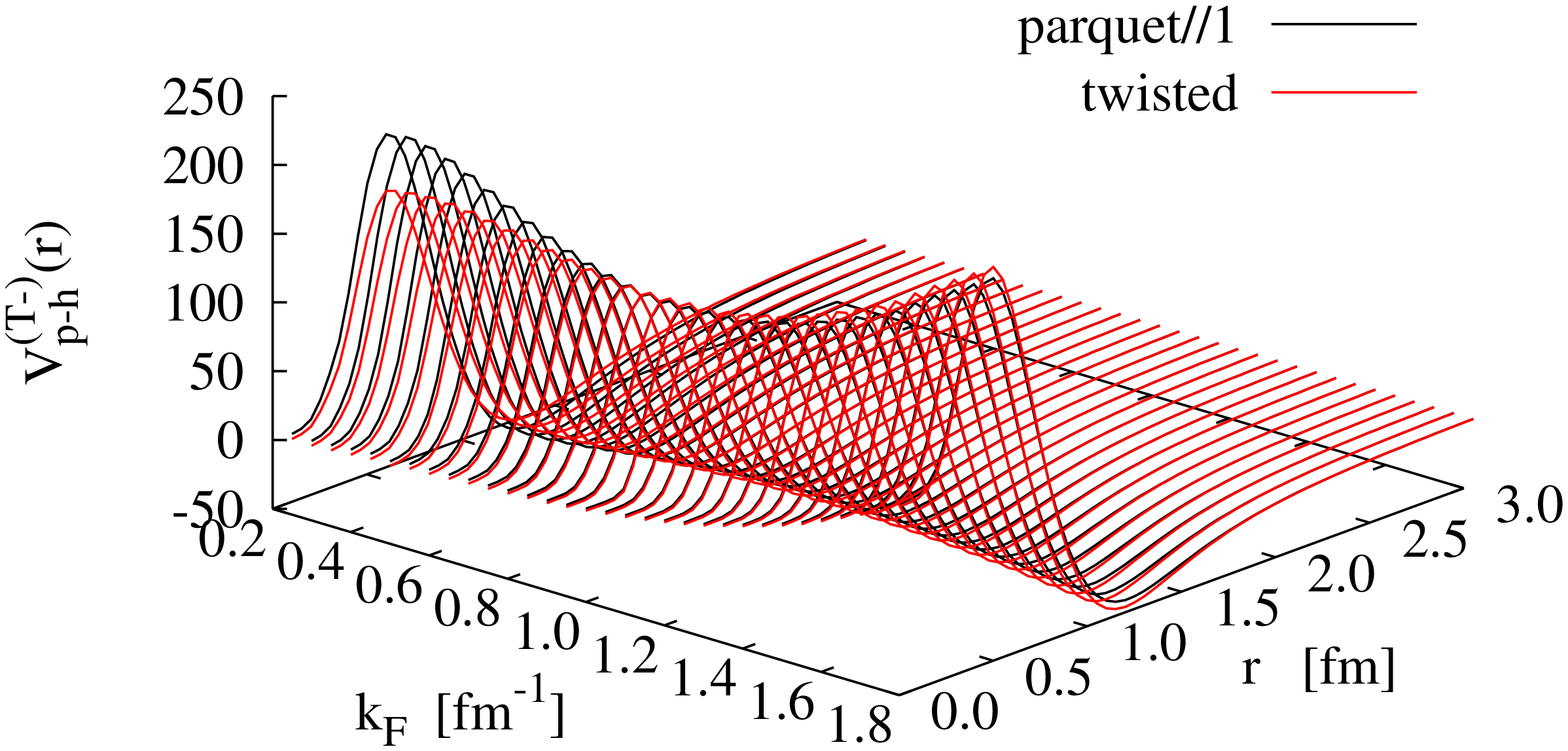}}
\label{fig:re_vph}
\caption{The figures show, for the Reid $v_6$
     interaction  the coordinate-space representation
  of the particle-hole interaction without (black lines) and
  with (red lines) the ``twisted chain'' corrections as a function
  of the Fermi wave number in the projector channels $S$, $T+$, and $T-$.}
\label{fig:Vph_3d}
  \end{figure}

Figs. \ref{fig:Vph_3d} shows the full particle-hole interaction in the
three projections $\{ S, T+, T-\}$ with and without $V_{\rm I}(r)$. Since
the $S$-channel effective interaction is the strongest -- this is
partly due to the strong nearest neighbor peak of the pair wave
function, see Fig. \ref{fig:Reid} -- the many-body effects are
comparatively weak {\em despite\/} the fact that $V_{\rm I}(r)$ is the
dominant effect. The total effect is much larger in the two triplet
channels $T+$ and $T-$ and can be as large as a factor of two at low
densities.

We conclude this section by remarking that the importance of
non-parquet diagrams is much less visible in momentum space, this is
basically caused by the fact that the long-wavelength limit of both
the particle-hole interaction and the induced interaction are
determined by Fermi-Liquid parameters which come out reasonably well
even in the ordinary FHNC-EL or parquet theory.
  
\section{Summary}
\label{sec:summary}

We have in this paper developed a procedure to go beyond
parquet-diagram calculations in a nuclear many-body Hamiltonian. The
essential aspect of that Hamiltonian is the state-dependence of the
interaction. We have analyzed the symmetrized operator product form of
the wave function of the type \eqref{eq:f_prodwave} and have come to
the conclusion that commutator corrections, which have so far been
ignored, can massively compromise the vailidity of low-order methods ,
can be very important in cases where the interactions in spin-singlet
and spin-triplet states are very different. The problem largely
removed in parquet theory that can be formulated in trms of physical
observables and has no need for introducing variational correlation
functions.

\begin{figure}
  \centerline{\includegraphics[width=0.7\columnwidth]{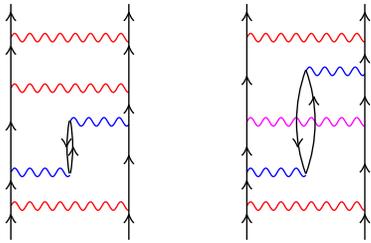}}
  \caption{(color online) The figure shows the essential processes are
    included in the ``twisted chain'' interaction correction. The red
    wavy lines are either spin-singlet or spin-triplet interactions,
    the magenta line may be either of the two, and the blue lines
    represent the induced interaction $\hat w_{\rm I}$.\label{fig:vsummary}}
  \end{figure}

The physical mechanism for why this is the case is made clear by
looking at the relevant processes from the point of view of
diagrammatic perturbation theory. The relevant mechanism is summarized
in figure \ref{fig:vsummary}. In the left diagram, a pair of particles
that enter the process in a specific (singlet or triplet) state will
always remain in that state. The red wavy lines therefore describe
interactions in the {\em same\/} channel. This is not changed by the
exchange of a spin-fluctuation despite the fact that the blue lines
may be singlet or triplet interactions.

In the right diagram, a spin is absorbed, transported through a
spin-fluctuation (described by the chain of two blue wavy lines), and
re-absorbed at a later time. Therefore, the magenta wavy line may be a
triplet interaction whereas the red lines are singlet interactions or
vice versa.  Evidently, this makes little difference if the
interactions are the same in spin-singlet and spin-triplet states. On
the other hand, there is no reason that this is a valid approximation
if the interactions are very different which is the case for modern
nucleon-nucleon interactions \cite{Reid68,AV18}.

On the technical side we have utilized techniques from both
variational Jastrow-Feenberg theory and perturbation theory.  The
analysis of the symmetrized operator product form of the variational
wave function has indicated the potential importance of commutator
corrections. The correspondence between Jastrow-Feenberg and Goldstone
diagrams has then revealed that these commutator corrections
correspond to Goldstone diagrams outside the parquet class, it also
suggested a way to calculate these corrections that would not be
immediately obvious from just looking at, for example, the third
diagram in Fig. \ref{fig:v5twist}.

To deal with this effect, we have utilized experience from both
variational and perturbation theory. We have used the correspondence
between Jastrow-Feenberg and Goldstone diagrams to conclude that these
processes are not described by parquet diagrams. The practical
implementation of these terms utilized again the view of variational
wave functions to identify approximations for those non-parquet
diagrams that would not be obvious from a purely perturbative point of
view.

From the analysis of the commutator diagrams one might have expected a
larger effect on the energetics of the system. The reader is reminded
that the argument applies only when the correlation functions
$f_\alpha(r)$ are determined by some low-order methods {\em and\/}
commutator corrections are included. We have shown in previous work
\cite{SpinTwist}, which is briefly outlined in the appendix, that this
effect can be drastic. FHNC-EL completely eliminates the need for
introducing correlation functions $f_\alpha(r)$ and is formulated
entirely in terms of the pair distribution function or the direct
correlation function $\Gamma_{\rm dd}(r)$, parquet theory never even
introduces such correlation functions. That way, the problem of
potentially divergent contributions never occurs which can otherwise
only be solved by omitting them.

The results have been described in Section \ref{sec:results}, there is
no need for repetition. The effect of the non-parquet contribution on
the short-ranged correlations and the effective interactions in the
spin-singlet channel at low densities is enhanced by the relatively
strong attraction. We have commented on this effect in earlier work
\cite{v3lett}.

The strong $S$-wave interaction has led to discussions of a potential
BCS-BEC crossover in low-density neutron matter
\cite{NuclBCSBEC,SchuckBCS2018}, our findings would suggest that
many-body effects can suppress such a crossover. It must be kept in
mind, however, that the repulsive interaction in the spin-triplet
channels is responsible for stabilizing neutron matter. A model system
of nucleons interacting in all channels with the $S$ wave interaction
would have a very low density spinodal point and be unstable at any
density that might be of interest for the structure of neutron stars.

We have shown here only the most essential results, namely effective
interactions which are input to calculations of pairing phenomena or
low-lying excitations. For recent review articles on pairing phenmena,
see Refs. \onlinecite{SchuckBCS2018} or
\onlinecite{SedrakianClarkBCSReview} and a collection of papers
describing recent research\cite{ECTSI}.

Similarly important is the response of neutron matter which has been
discussed over the years \cite{Wam93,Benhar2009,Lovato2013}.  A
particular promising route appears to be the extension of the pair
excitation theory \cite{eomIII,2p2h} to nuclear cases which have
provided a quantitative understanding of the full dynamic structure
function of quantum fluids
\cite{Nature_2p2h,skwpress,LichtenDiss}. The method may be understood
as a correlated version of what is called in nuclear physics ``second
RPA\cite{SRPA83,SRPA87}'', being built on a correlated ground state
instead of a model state of single particle wave functions, the
approach avoids the usual problems caused by strong, short-ranged
correlations. Another important further extension of our methods is,
of course, the inclusion of spin-orbit forces which are of
quantitative importance \cite{Spin-OrbitPolls2005}.  Work in this
direction is in progress.

\appendix
  \section{A simple example}

We review in this section a somewhat simpler case where the effect of
the symmetrization can be studied explicitly. Consider a fictitious
system of bosons with spins \cite{SpinTwist}. We keep only the $\hat
O_c = \1$ and $\hat O_3(i,j) = \sij{i}{j}$.  In that case, the
cluster expansions can be simplified by assuming a symmetrized
operator product for the {\em square\/} of the wave function,
\begin{equation}
\Psi_0^2 = {\cal S}\biggl[\prod_{i<j}(f_c^2(r_{ij}) + f^2_\sigma(r_{ij})\sigma_i\cdot
  \sigma_j)\biggr]\,.\label{eq:psisquared}
\end{equation}
The distribution functions then have the general form
\begin{align}
g_c(r) &= f^2_c(r)F_{cc}(r) + f^2_\sigma(r)F_{c\sigma}(r)\nonumber\\
g_\sigma(r) &= f^2_c(r)F_{\sigma c}(r) + f^2_\sigma(r) F_{\sigma\sigma}(r)\,.
\end{align}
where the $F_{ij}$ are multidimensional integrals involving
$h_c(r_{ij}) = f^2_c(r_{ij})-1$ and $h_\sigma(r_{ij}) = f^2_\sigma(r_{ij})$. 
If one ignores all commutators, a set of HNC equations can be derived in much
the same way as for spin-independent correlations. The coefficient functions
$F_{\alpha\beta}(r)$ become
\begin{align}
  F_{cc}(r) &= \frac{1}{4}\left[3e^{N_\sigma(r)} + e^{-3N_\sigma(r)}\right]
  e^{N_c(r)}\nonumber\\
  F_{c\sigma}(r) &= \frac{3}{4}\left[e^{N_{\sigma}(r)} - e^{-3N_{\sigma}(r)}\right]
  e^{N_c(r)}= 3F_{\sigma c}(r) \nonumber\\
  F_{\sigma\sigma}(r) &= \frac{1}{4}\left[e^{N_{\sigma}(r)} + 3 e^{-3N_{\sigma}(r)}\right]e^{N_c(r)}\label{eq:Funtwist}
\end{align}
where the $N_{c,\sigma}(r)$ are the sums of chain diagrams. Defining
the sets of non-nodal diagrams
\begin{equation}
  X_{c}(r)=  g_{c}(r)-1- N_{c}(r)\,,\qquad
  X_{\sigma}(r)=  g_{\sigma}(r)- N_{\sigma}(r)
\end{equation}
the nodal diagrams  $N_{c,\sigma}(r)$ are given in momentum space,
\begin{equation}
  \tilde N_{c,\sigma}(k) =\tilde X_{c,\sigma}^2(k)/(1-\tilde X_{c,\sigma}(k))
\end{equation}

In the next step, the parallel connections of all possible chains are
symmetrized with the appropriate combinatorial factors. One then
obtains a different set of coupling coefficients
\cite{Lagaris85,SpinTwist}
\begin{align}
  F_{cc}(r) &= \biggl[\cosh(N_\sigma(r)) + N_\sigma(r)\sinh(N_\sigma(r))\biggr]
  e^{N_c(r)}\nonumber\\
  F_{c\sigma}(r) &= \biggl[2\sinh(N_\sigma(r)) + N_\sigma(r)\cosh(N_\sigma(r))
    \biggr]
  e^{N_c(r)}\nonumber\\
  &= 3 F_{\sigma c}(r)\nonumber\\
  F_{\sigma\sigma}(r) &= \biggl[\cosh(N_\sigma(r)) + \frac{1}{3}
    N_\sigma(r)\sinh(N_\sigma(r))\biggr]e^{N_c(r)}.\label{eq:Ftwist}
\end{align}
Eqs. \eqref{eq:Funtwist} and \eqref{eq:Ftwist} look, at the first glance,
innocuous. To demonstrate our point we rewrite the pair-distribution
functions in the singlet and triplet channels,
\begin{equation}
  g(r) = g_s(r) \hat P_s + g_t(r) \hat P_t
\end{equation}
where
\begin{equation}
  g_s(r) = g_c(r)-3g_\sigma(r),\qquad g_t(r) = g_c(r)+g_\sigma(r)
\end{equation}
are the
distribution functions and nodal quantities in these
channels. In this representation we have, for the {\em
  unsymmetrized\/} version \eqref{eq:Funtwist}
\begin{equation}
  g_{s,t}(r) = f_{s,t}^2(r)e^{N_{s,t}(r)}\,,
\end{equation}
\ie the distribution functions are indeed proportional to the
correlation functions in the spin-singlet and spin-triplet channels.
On the other hand, such a simple relationship can not be derived if
the simplest non-trivial commutators are included as in
Eqs. \eqref{eq:Ftwist}. The pair distribution functions $g_{s,t}(r)$
are combinations of $f_{s}^2(r)$ and $f_{t}^2(r)$ whose detailed
structure is not illuminating.

\begin{acknowledgments}
 Encouragement for this work was derived from a workshop on {\em
   Nuclear Many-Body Theories: Beyond the mean field approaches\/} at
 the Asia Pacific Center for Theoretical Physics in Pohang, South
 Korea, in July 2019. One of us (JW) thanks the Austrian Marshall Plan
 Foundation for support during the summer 2018 and Robert Zillich for
 discussions.
\end{acknowledgments}

%


\end{document}